# Synergy cycles in the Norwegian innovation system:

# The relation between synergy and cycle values

Inga Ivanova[1], Øivind Strand[2], & Loet Leydesdorff[3]

**Abstract**

The knowledge base of an economy measured in terms of Triple-Helix relations can be analyzed in terms of mutual information among geographical, sectorial, and size distributions of firms as dimensions of the probabilistic entropy. The resulting synergy values of a TH system provide static snapshots. In this study, we add the time dimension and analyze the synergy dynamics using the Norwegian innovation system as an example. The synergy among the three dimensions can be mapped as a set of partial time series and spectrally analyzed. The results suggest that the synergy at the level of both the country and its 19 counties shows non-chaotic oscillatory behavior and resonates in a set of natural frequencies. That is, synergy surges and drops are non-random and can be analyzed and predicted. There is a proportional dependence between the amplitudes of oscillations and synergy values and an inverse proportional dependency between the oscillation frequencies' relative inputs and synergy values. This analysis of the data informs us that one can expect frequency-related synergy-volatility growth in relation to the synergy value and a shift in the synergy volatility towards the long-term fluctuations with the synergy growth.

## 1. Introduction

Multi-dimensional systems of various types, such as social or biological, can be considered eco-systems, that can flourish if uncertainty in the relations among constituent parts is reduced (Ulanowicz, 1986). The Triple Helix (TH) model of university-industry-government relations can serve as a specific example of such systems. Mutual information in three or more dimensions can be considered a reduction of uncertainty at the system level or a measure of synergy, and can be expressed in terms of bits of information using the Shannon-formulas (Abramson, 1963; Theil, 1972; Leydesdorff, 1995).

[1] School of Business and Public Administration, Far Eastern Federal University, 8, Sukhanova St., Vladivostok 690990, Russia; inga.iva@mail.ru
[2] Aalesund University College, Department of International Marketing, PO Box 1517, 6025 Aalesund, Norway; +47 70 16 12 00; ost@hials.no
[3] University of Amsterdam, Amsterdam School of Communication Research (ASCoR), PO Box 15793, 1001 NG Amsterdam, The Netherlands; loet@leydesdorff.net

The synergy of a TH system can be measured as reduction of uncertainty using mutual information among the three dimensions of firm sizes, the technological knowledge bases of firms, and geographical locations. Mutual information can be expressed in bits of information using the formalisms of Shannon's information theory. One should note that the three dimensions refer to different institutional actors and are functionally differentiated. Because of the additive character of entropy the system of university-industry-government relations can graphically be displayed in the form of a Venn diagram with each surface area corresponding to the expected information content (Fig. 1).

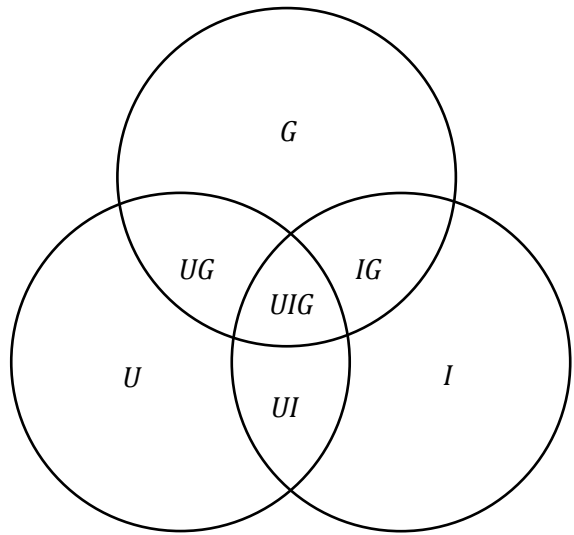

Fig. 1 Venn diagram of university (*U*) –industry (*I*) –government (*G*) relations. The central overlapping part *UIG* corresponds to the three-lateral mutual information.

The problem in applying Shannon formalism to three-lateral and higher-order dimensional interactions is that mutual information is then a signed information measure (Yeung 2008, Leydesdorff 2010). A negative information measure cannot comply with Shannon's definition of information (Krippendorff 2009a, b). This contradiction can be solved by considering mutual information as different from mutual redundancy (Leydesdorff & Ivanova, 2014). In the three-dimensional case, however, mutual information is equal to mutual redundancy and, thus, can be considered a Triple-Helix indicator of synergy in university-industry-government relations (Leydesdorff *et al*, 2014).

A number of studies have been devoted to measuring synergy across different countries and regions, such as the Netherlands (Leydesdorff, Dolfsma, & Van der Panne, 2006), Germany (Leydesdorff & Fritsch, 2006), Hungary (Lengyel & Leydesdorff, 2011), Norway (Strand & Leydesdorff, 2013), Sweden (Leydesdorff & Strand, 2012), West Africa (Mêgnigbêto, 2013), China (Leydesdorff & Zhou, 2014), and Russia (Leydesdorff, Perevodchikov, & Uvarov, in press). One obtains maps of synergy distribution across the territory. However, having only these synergy "snapshots", one is unable to answer a series of questions, such as what is the temporal character of synergy evolution, and does the synergy value affect its temporal evolution? Note that a TH cannot be static (Etzkowitz & Leydesdorff, 2000). Rather it is an ever-evolving system, and therefore one can expect that the synergy in this system also evolves with the passage of time.

The core research questions of the present paper regarding temporal synergy evolution are as follows: how does the synergy evolve (e.g., is there a trend-like, chaotic, oscillatory, or some other functional dependency)? Do synergy values affect the temporal evolution (i.e. is there a difference in synergy evolution between high and low synergy). And can we provide numerical indicators of synergy evolution? Answering these questions may shed light on the control mechanisms of a system and provide tools for exploring multi-dimensional systems of this type in different areas.

In this study, we analyze the temporal dynamics of mutual information in the Norwegian innovation system as an example. The choice of the Norwegian innovation system is guided by the ready availability of data. However, the method is generic and can be applied to any data for time series that fulfill the criterion of possessing three (or more) different dimensions. The paper is structured as follows. Section 2 describes the method. The results are presented in Section 3 and discussed in Section 4. Finally some conclusions and policy implications are formulated in Section 5.

## 2. Methods and data

*2.1 Methods*

The synergy of interaction between two actors can be numerically evaluated using the formalisms of Shannon's information theory by measuring mutual information as the reduction of uncertainty. In the case of three interacting dimensions, the mutual (configuration) information $T_\Sigma$ can be defined by analogy with mutual information in two dimensions, as follows (Abramson, 1963; McGill, 1954):

$$T_\Sigma = H_1 + H_2 + H_3 - H_{12} - H_{13} - H_{23} + H_{123} \qquad (1)$$

Here, $H_i$, $H_{ij}$, $H_{ijk}$ denote probabilistic entropy measures in one, two, and three dimensions:

$$H_i = -\sum_i p_i \log_2 p_i$$

$$H_{ij} = -\sum_{ij} p_{ij} \log_2 p_{ij} \qquad (2)$$

$$H_{ijk} = -\sum_{ijk} p_{ijk} \log_2 p_{ijk}$$

The values of *p* represent the probabilities, which can be defined as the ratio of the corresponding frequency distributions:

$$p_i = {n_i}/{N}; \; p_{ij} = {n_{ij}}/{N}; \; p_{ijk} = {n_{ijk}}/{N} \qquad (3)$$

$N$ is the total number of events, and $n_i$, $n_{ij}$, $n_{ijk}$ denote the numbers of events relevant in subdivisions. For example, if $N$ is the total number of firms, $n_{ijk}$ is the number of firms in the $i$-th county, the $j$-th organizational level (defined by the number of staff employed), and the $k$-th technology group. Then $n_i$ and $n_{ij}$ can be calculated as follows:

$$n_i = \sum_{jk} n_{ijk}; \qquad n_{ij} = \sum_k n_{ijk}$$

A set of $L$ mutual information values for a certain time period, considered as a finite time signal, can be spectrally analyzed with the help of the discrete Fourier transform (Kester, 2000):

$$T_\Sigma = \sum_{l=0}^{L/2} F_l(w) \qquad (4)$$

Here:

$$F_0 = A; \; F_l(w) = B_l \cos(2\pi l w/L) + D_l \sin(2\pi l w/L) \qquad (5)$$

The Fourier decomposition by itself cannot provide us with information regarding synergy evolution except the values of the spectral coefficients $A$, $B_i$, and $D_i$. Because the aggregate (country-related) synergy $T_\Sigma$ is determined by additive entropy measures (Eq. (1)), it can also be decomposed as a sum of partial (county-related) synergies $T_1, \dots T_n$:[*]

$$T_\Sigma = T_1 + T_2 + \cdots T_n \qquad (6)$$

So that each partial synergy can be written in the same form as Eq. (4):

---

[*] This decomposition is different from that used in our previous studies (e.g., Leydesdorff & Strand, 2013; Strand & Leydesdorff, 2013).

$$T_1 = \sum_{l=0}^{L/2} f_{1l}(w)$$

$$T_2 = \sum_{l=0}^{L/2} f_{2l}(w)$$

(7)

...

$$T_L = \sum_{l=0}^{L/2} f_{1L}(w)$$

Here:

$$f_{0l} = a_{0l};\ f_{nl}(w) = b_{nl}\cos(2\pi lw/L) + d_{nl}\sin(2\pi lw/L)$$

After substituting Eqs. (4) and (7) into (6) and re-grouping the terms, one obtains:

$$F_l(w) = f_{1l}(w) + f_{2l}(w) + .. + f_{nl}(w) \qquad (8)$$

Leydesdorff and Ivanova (2014) showed that mutual information in three dimensions is equal to mutual redundancy ($T_{123} = R_{123}$). Aggregated redundancy can equally be decomposed as a sum of partial redundancies, corresponding to the geographical, structural, or technological dimensions of the innovation system under study. Mutual redundancy changes over time, so that one can write:

$$R_{123}(t) = R_1(t) + R_2(t) + \cdots + R_n(t) \qquad (9)$$

In another context, Ivanova & Leydesdorff (2014 b) expressed the redundancy that can be obtained as follows (*i*= 1, 2 … *n*):

$$R_i = a'_i + b'_i \cos(r_i t) + d'_i \cos(r_i t) \qquad (10)$$

The oscillating function in Eq. (10) can be considered a natural frequency of the TH system. This natural frequency is far from fitting observed redundancy values for $R_{123}$. However, real data for the definite time interval can be fit with the help of the discrete Fourier transform, comprising a finite set of frequencies. Each frequency in the set composing Eq. (9) can be considered a natural frequency of the TH system:

$$R_{123} = A + \sum_{k=1}^{n}(B_k \cos(kt) + D_k \sin(kt)) \qquad (11)$$

Comparing Eq. (11) with Eq. (10) one can approximate the empirical data for three-dimensional redundancy $R_{123}$ as a sum of partial redundancies $R_i$ corresponding to frequencies that are multiples of the basic frequency: *w, 2w, 3w* … etc.

$$R_{123} = R_1 + R_2 + \cdots + R_n \qquad (12)$$

In other words, a TH system can be represented as a string resonating in a set of natural frequencies with different amplitudes. Frequency-related amplitudes, which can be defined as modules of the corresponding Fourier coefficients, can be considered the spectral structure of the TH system. Absolute values of the Fourier-series coefficients $C_k$ can be defined as follows

$$C_l = \sqrt{(B_l^2 + D_l^2)} \qquad (13)$$

These coefficients determine the relative contributions of the harmonic functions with corresponding frequencies to the aggregate redundancy ($R_{123}$ in Eq. (11)).

*2.2  Data*

Norwegian establishment data were retrieved from the database of Statistics Norway at https://www.ssb.no/statistikkbanken/selecttable/hovedtabellHjem.asp?KortNavnWeb=bedrifter&CMSSubjectArea=virksomheter-foretak-og-regnskap&PLanguage=1&checked=true. The data include time series of Norwegian companies for the period 2002-2014, and encompass approximately 400,000 firms per year. The data include the number of establishments in the three relevant dimensions: geographical (*G*), organizational (*O*), and technological (*T*).

Nineteen counties are distinguished in the geographical dimension. In the organizational dimension, establishments are subdivided with reference to different numbers of employees by eight groups: no-one employed; 1-4 employees; 5-9 employees; 10-19 employees; 20-49 employees; 50-99 employees; 100-249 employees; and 250 or more employees. The number of employees can be expected to correlate with the establishment's organizational structure.

The technological dimension indicates domains of economic activity. The data for the period 2002-2008 were organized according to the NACE Rev. 1.1 classification, and the data for the period 2009-2014 were organized according to the NACE Rev. 2 classification. Some of the criteria for construction of the new classification, were reviewed: but there is no one-to-one correspondence between NACE Rev. 1.1 (with 17 sections and 62 divisions) and NACE Rev. 2 (with 21 sections and 88 divisions) (EUROSTAT a). To correctly merge the NACE Rev. 1.1 and NACE Rev. 2 data one has to turn to a higher level of aggregation (Appendix B) containing 10 classes (EUROSTAT b).

## 3. Data analysis

*3.1 Descriptive statistics*

Country synergy is decomposed as a sum of the synergies at the county level in accordance with Eq. (6). The results of the calculations for the period 2002-2014 years (in mbits of information) are shown in Figs 2-6.

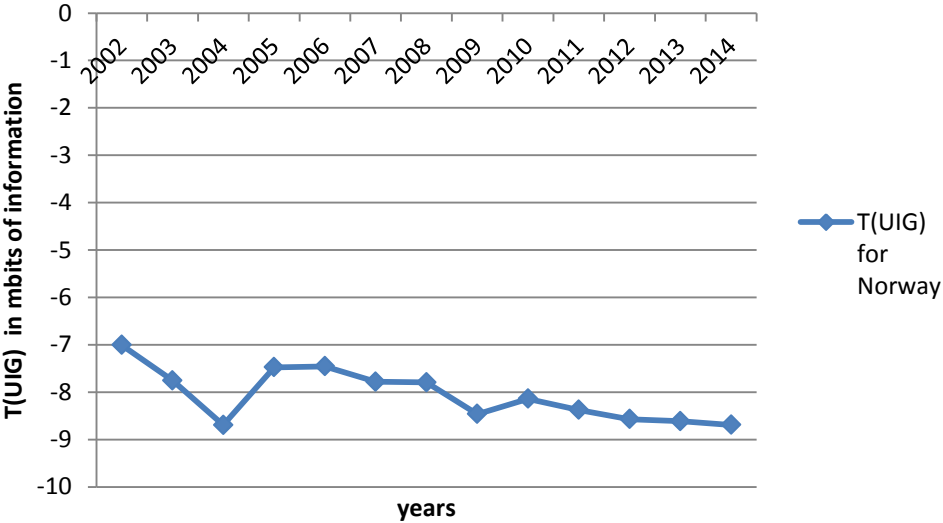

Fig. 2 Summary of the development of Norwegian synergy for the period 2002-2014 (in mbits of information)

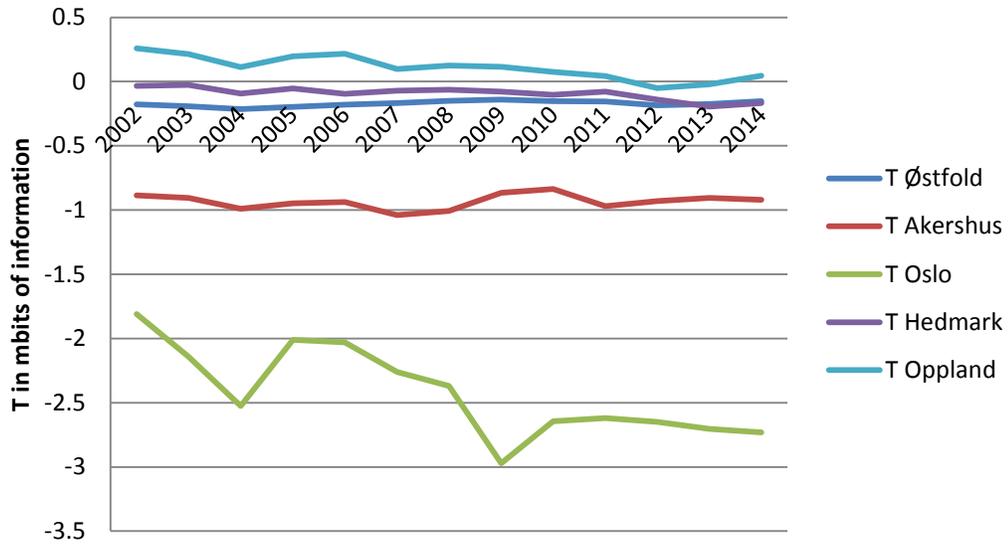

Fig. 3 Partial ternary synergy for Østfold, Akershus, Oslo, Hedmark, and Oppland, for the period 2002-2014 (in mbits of information)

The only county with a positive value for the mutual information in the three dimensions is Oppland. This may indicate the absence of synergy in *U-I-G* interactions in Oppland. However, a weak trend towards synergy can be detected. The trend for the capital Oslo shows a pattern similar to that for the nation. An increase in synergy, which is manifested by a more negative value of *T*, can be detected. In Strand & Leydesdorff (2013), the synergy calculations were based on municipal data, resulting in a singularity in the capital of the country (Oslo). In this paper, the calculations are based on the contributions of the counties to the national level, allowing the contribution of the capital to be specified.

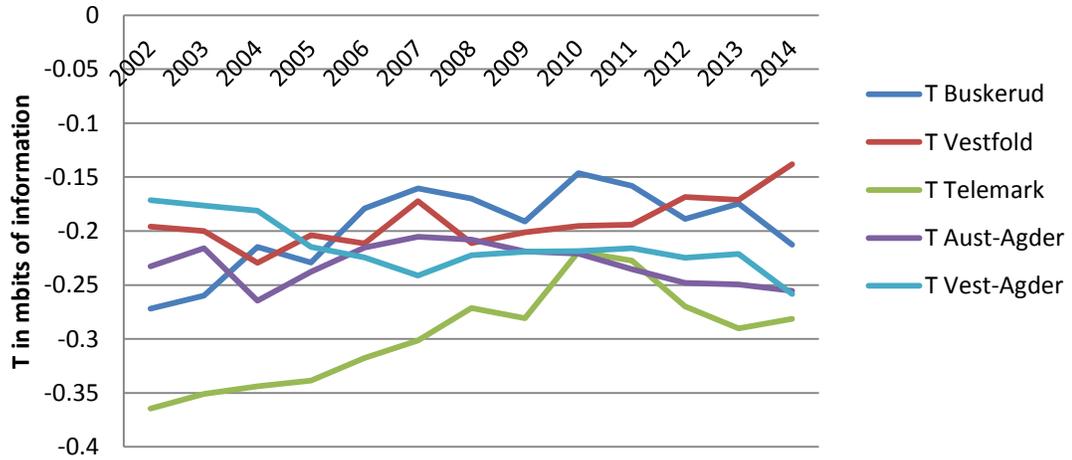

Fig. 4 Partial ternary synergy for Buskerud, Vestfold, Telemark, Aust-Agder, and Vest-Agder, for the period 2002-2014 (in mbits of information)

The industrialized counties of Telemark, Buskerud, and Vestfold show a pattern with decreasing synergy over time. The two Agder counties show an opposite development to the others with an increase in synergy.

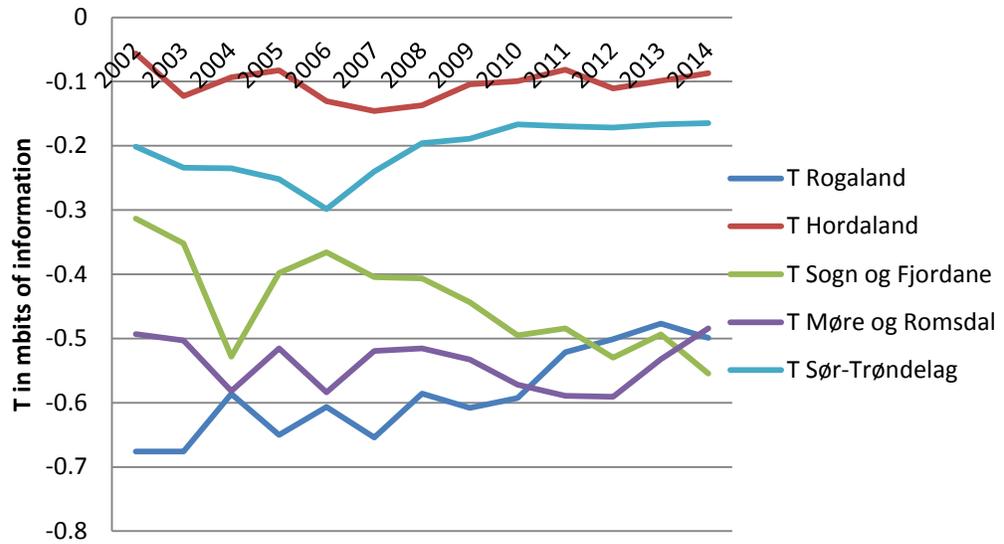

Fig. 5 Partial ternary synergy for Rogaland, Hordaland, Sogn og Fjordane, Møre og Romsdal, and Sør-Trøndelag, for the period 2002-2014 (in mbits of information)

Rogaland which is dominated by the oil industry, shows a decreasing trend, but the magnitude of the synergy still exceeds its neighbor Hordaland. The small counties of Sogn and Fjordane show a trend with increased synergy over time.

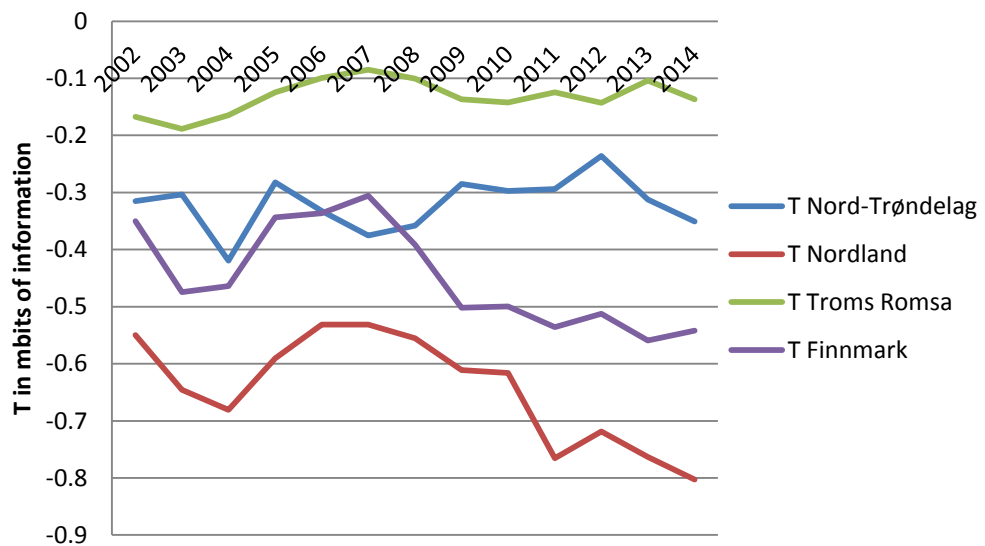

Fig. 6 Partial ternary synergy for Nord-Trøndelag, Nordland, Troms Romsa, and Finnmark, for the period 2002-2014 (in mbits of information)

Among the northern counties, Nordland and Finnmark show a development with an increase in synergy. Fluctuations in synergy data can be interpreted as synergy cycles. Like economic cycles they may indicate some endogenous characteristics of an innovation system such as cyclic oscillations of the market system (Morgan, 1991). An alternative to considering the fluctuations as cycles would be to consider them a result of noise in the data; we clarify this point in the next section.

*3,2 Transmission power and efficiency*

Having the transmission time series we calculated the transmission power time series for Norway as a whole and separately for constituent counties according the following formula (Mêgnigbêto, 2014, p. 287):

$$\tau = \begin{cases} \tau_1 = \frac{T_{GOT}}{H_{GOT}-H_G-H_O-H_T} & if\ T_{GOT} < 0 \\ \tau_2 = \frac{T_{GOT}}{H_{GOT}} & if\ T_{GOT} > 0 \\ 0 & if\ T_{GOT} = 0 \end{cases} \quad (14)$$

The transmission power was designed to measure the efficiency of the mutual information. While the transmission defines the total amount of configurational information, the transmission power represents the share of the synergy actually produced in the system relative to its size. For positive transmission values, it is simply the ratio of overlapping surface area in the Venn diagram to the whole surface area of the figure (Fig. 1). Mêgnigbêto (2014, p.290) argued that "… with such indicators, a same system may be compared over time; different systems may also be compared". Figs. 7-11 present the graphs of the transmission power.

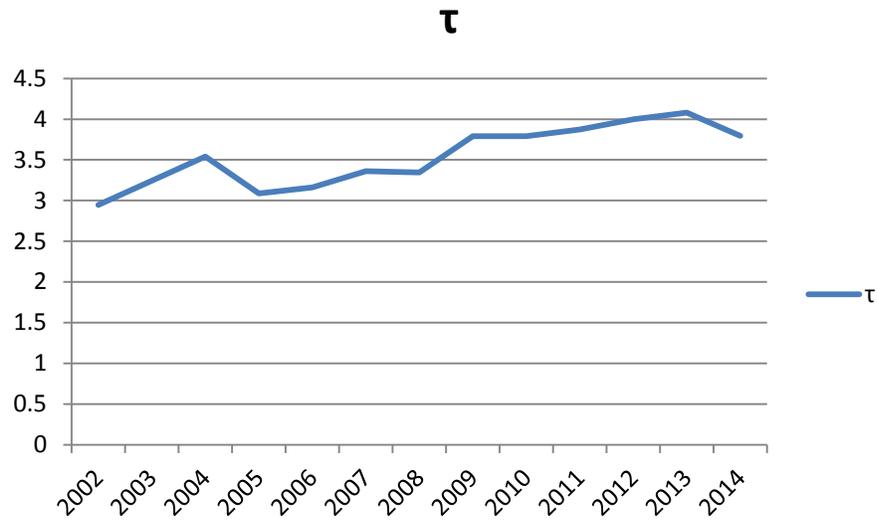

Fig. 7 Summary Norway transmission power $\tau$ (in relative units*100) for the period 2002-2014

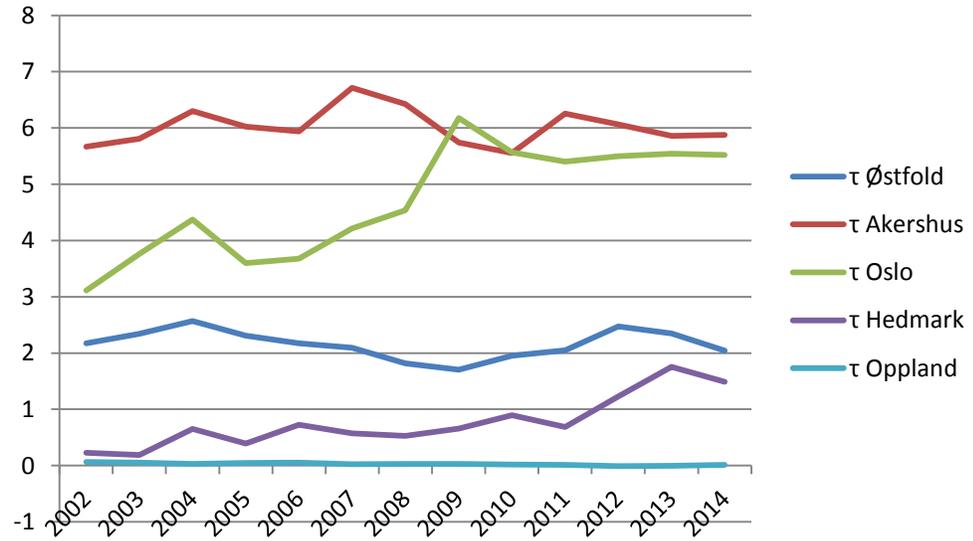

Fig. 8 Transmission power $\tau$ for Østfold, Akershus, Oslo, Hedmark, and Oppland, (in relative units*100) for the period 2002-2014

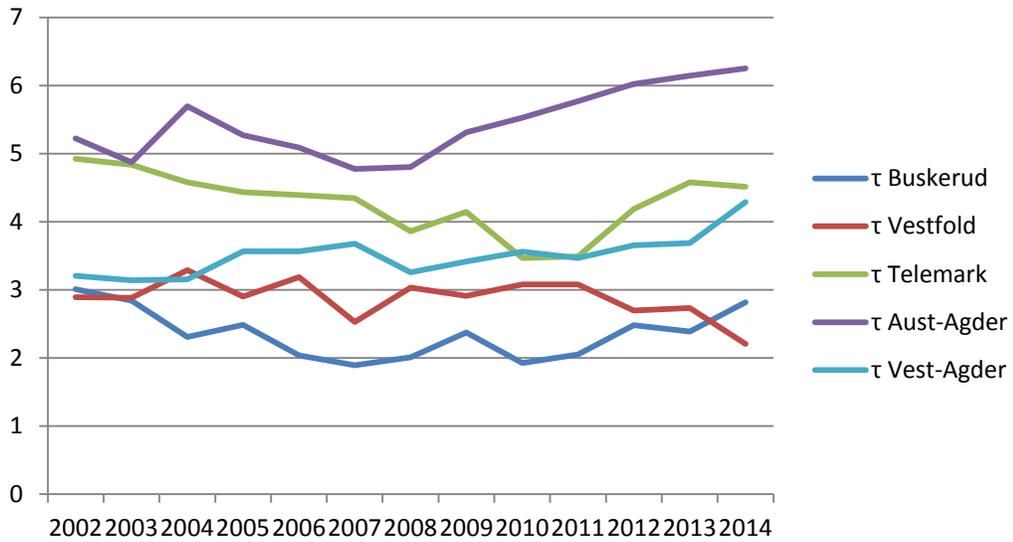

Fig. 9 Transmission power τ for Buskerud, Vestfold, Telemark, Aust-Agder, and Vest-Agder, (in relative units*100) for the period 2002-2014

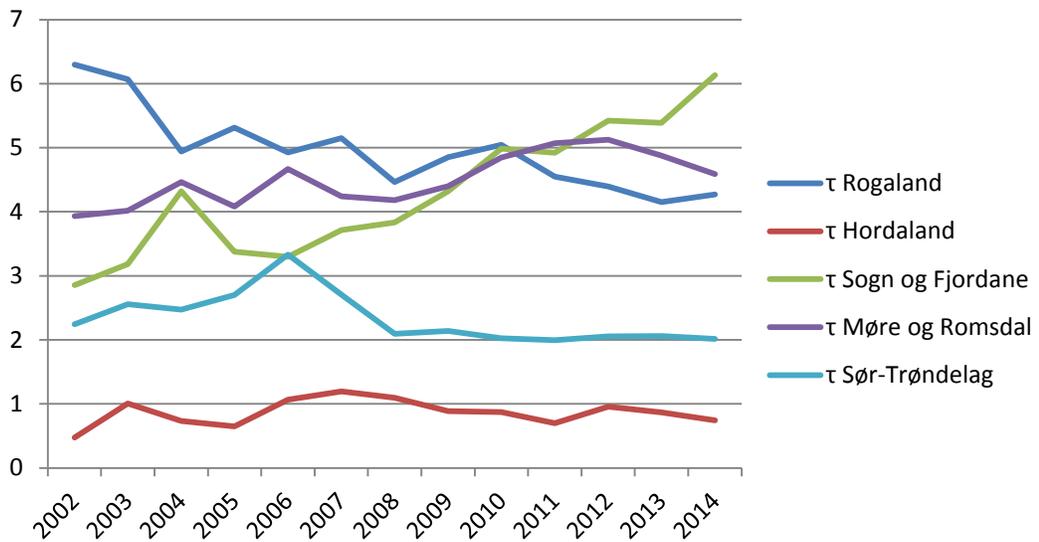

Fig. 10 Transmission power τ for Rogaland, Hordaland, Sogn og Fjordane, Møre og Romsdal, and Sør-Trøndelag, (in relative units*100) for the period 2002-2014

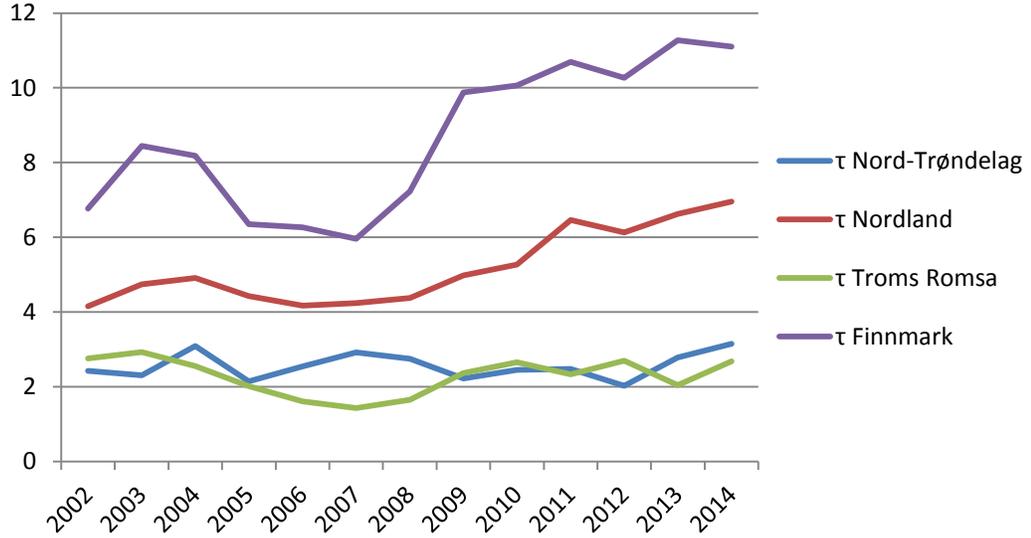

Fig.11 Transmission power τ for Nord-Trøndelag, Nordland, Troms Romsa, and Finnmark, (in relative units*100) for the period 2002-2014

Comparing the national level transmission power in Fig. 7 with the synergy in Fig. 2 shows increased transmission power and increased synergy over time. At the county level, the same patterns are most pronounced in Oslo, the Agder counties, Sogn og Fjordane, and Nordland og Finmark.

We compared the percentage of the average efficiency deviation $K = \frac{\tau_{iav} - \bar{\tau}_{iav}}{\bar{\tau}_{iav}} * 100\%$, where $\tau_{iav}$ is the efficiency for the *i*-th county averaged over the period 2002-2014; $\bar{\tau}_{iav}$ is the summary average efficiency averaged over all of the counties (Fig. 11), and the percentage of average synergy deviation $P = \frac{T_{iav} - \bar{T}_{iav}}{\bar{T}_{iav}} * 100\%$, where $T_{iav}$ is the synergy for *i*-th county averaged over the period 2002-2014; and $\bar{T}_{iav}$ is the summary average synergy averaged over all of the counties (Fig.12). Efficiency is above the country average in Akershus (τ 02), Oslo (τ 03), Aust-Agder (τ 09), Rogaland (τ 11), Sogn og Fjordane (τ 14), Møre og Romsdal (τ 15), and Nordland (τ 18), and is extremely high in Finnmark (τ 20). One can observe that the efficiency and synergy peaks do not coincide. That is counties with the highest synergy values cannot always be considered the most efficient. This may indicate that the increase in synergy is caused by increased transmission power.

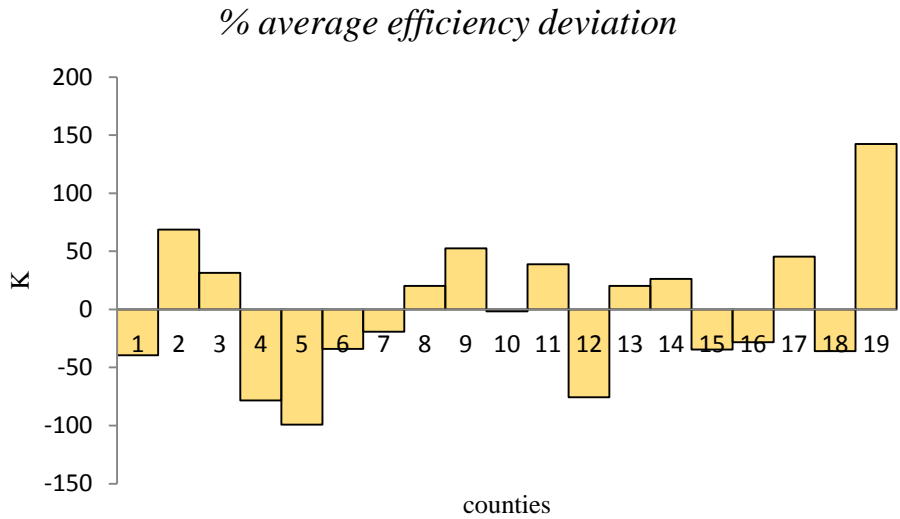

Fig. 12 Percentage of average efficiency deviation for 19 Norwegian counties for the period 2002-2014 (in percent)

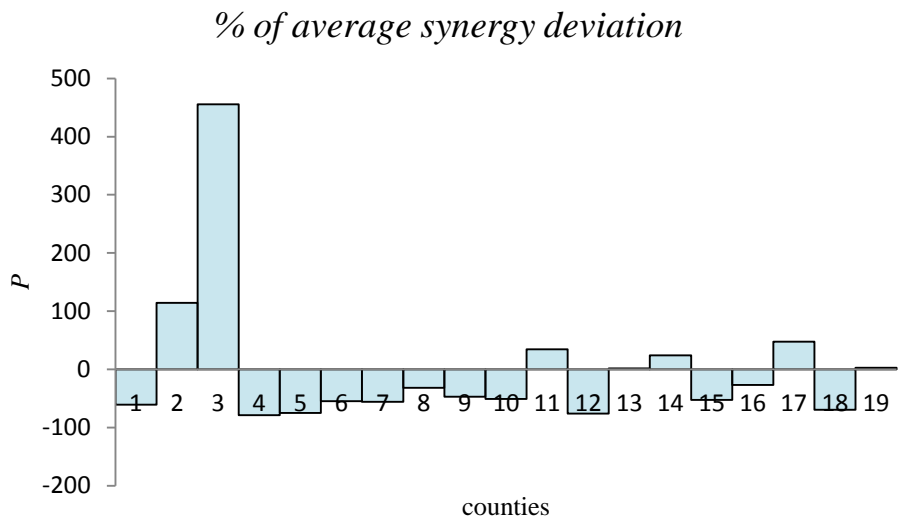

Fig. 13 Percentage of average synergy deviation for 19 Norwegian counties for the period 2002-2014 (in percent)

As a next step, we analyzed aggregate redundancy time series with the help of the discrete Fourier transform in accordance with Eq. (4). The inputs of different frequency modes to Norway's synergy ($w$, $2w$, $3w$, $4w$, $5w$, $6w$), calculated according to Eq. (14), are shown in Fig. 14.

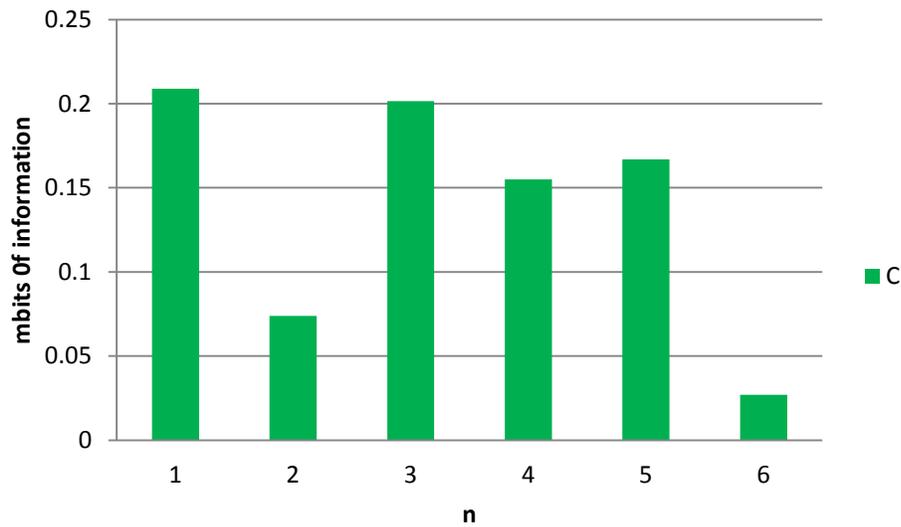

Fig. 14 Modules of Fourier series coefficients C versus frequency for summary ternary synergy (in mbits of information)

Each of the county-related synergies can be mapped as fluctuations around an average value. Thus, the average values can be taken as the first terms in the corresponding Fourier decomposition describing non-fluctuating terms ($f_{0i}$ in Eq. (7)). These average values form the synergy line specter.

Having calculated the modules of the Fourier series coefficients, which are the measures of different frequency modes, as well as the line specter synergy values we can map these modules versus synergy values to obtain corresponding functional dependencies. Because we address the real-number data (for the period 2002-2014), then, due to the symmetry of DFT coefficients, only half the number of input data with different frequency components (the first six) can be discerned.

In Fig. 15 synergies (in mbits of information) are plotted versus frequency amplitudes (in mbits of information). It can be seen from the figure that observed synergies can be fitted by polynomial approximation. The coefficients of determination $R^2$ of the approximation curve for all of the graphs, except for *2w*, are relatively high. The coefficients at the first term, defining the long-run speed of the corresponding frequency relative to the contribution increase, decreases from the low-frequency end to the high-frequency end of the specter. Extreme synergy and Fourier coefficient values are also found for Oslo. The equations in the upper parts of graphs denote the explicit form of the polynomial approximation

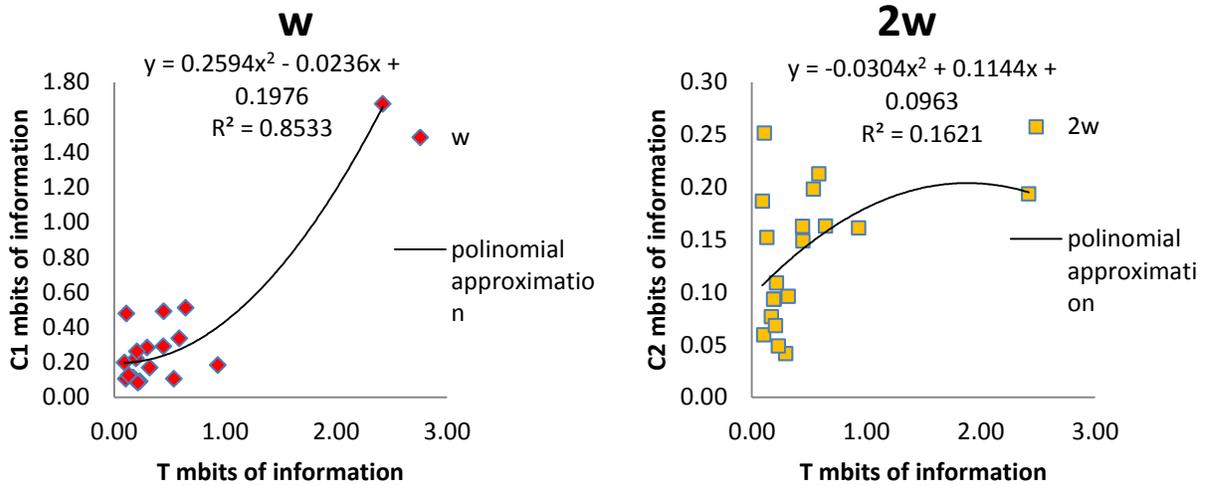

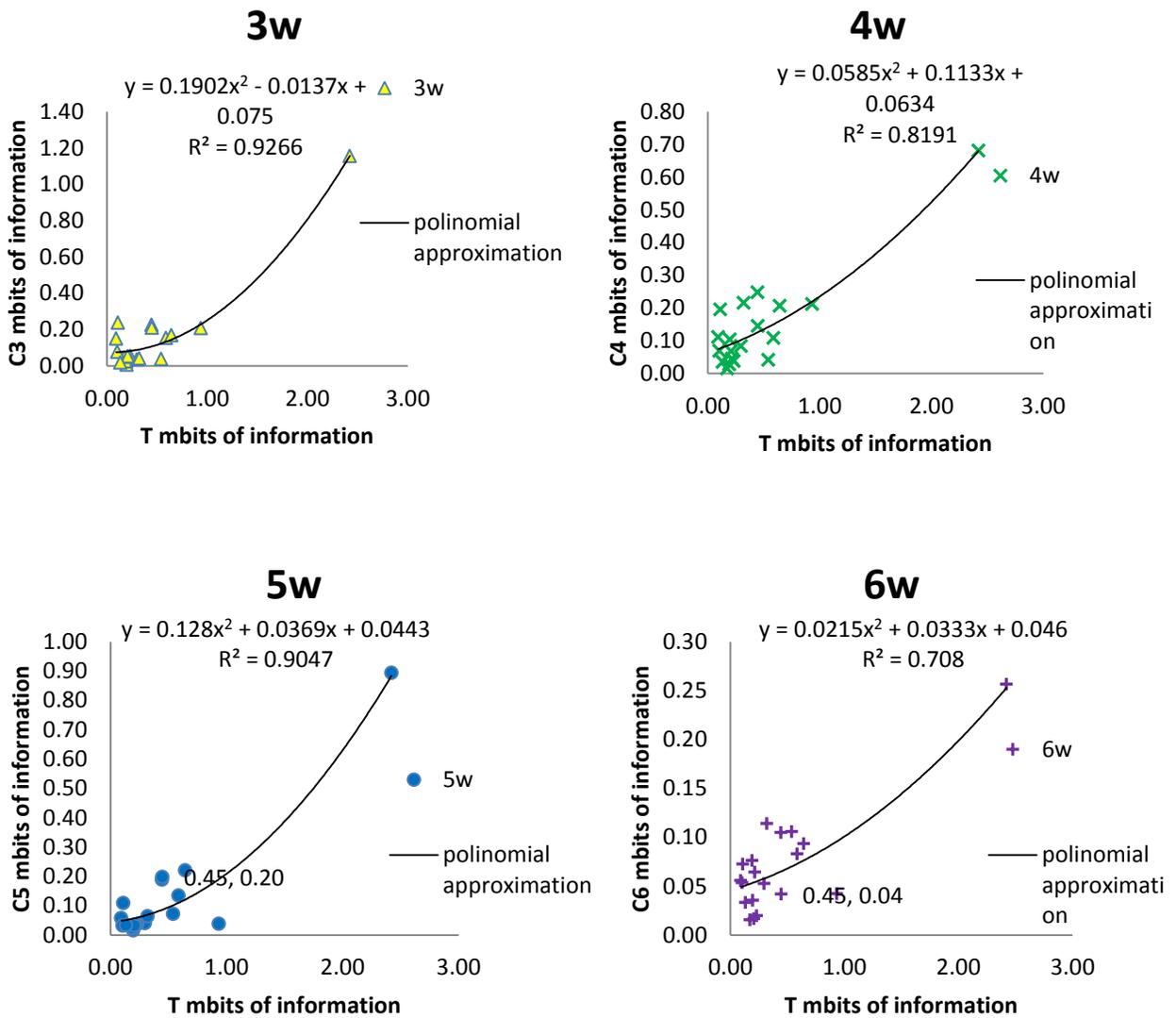

Fig. 15 Observed average synergy absolute values (in mbits of information) versus Fourier coefficients absolute values (in mbits of information) for the first six frequencies.

The degree of synergy fluctuation randomness can be evaluated using R/S analysis (Hurst, 1951). The standard algorithm and the calculation results are presented in the Appendix A. The Hurst rescaled range statistical measure H values in the range $0.5 < H < 1$ indicate a persistent or trend-like behavior described by monotone function. $H = 0.5$ corresponds to a completely chaotic time series behavior, like that of Brownian noise. Values in the range $0 < H < 0.5$ indicate anti-persistent or oscillating behavior.

The obtained Hurst exponent value, in our case $H = 0.065$, is well below 0.5 indicating a strongly expressed oscillating time series behavior. That is, the system-generated synergy evolves over time as non-chaotic cycles (similar to long-term and business cycles).

## Summary and Conclusions

In the presented methodological approach of numerically evaluating temporal synergy evolution in a three-dimensional functionally differentiated system, we relied on the 'coherent input' of three non-intercepting technics: R/S analysis, DFT, and geographical synergy decomposition. Briefly summarizing the results obtained from the study of the Norwegian innovation system, we can conclude that the synergy time series exibits cyclic structure of a non-random nature. This is important from the perspective that synergy oscillations can be caused, in part, by system-inherent factors, and, in part, by outer systemic factors. This feature should be taken into consideration by policy makers when developing related policies in innovation or other relevant spheres.

From the conceptual viewpoint, the synergy in the TH systems can be presented as a self-resonating set of the system's harmonic partials which are the same at the country and county levels. An unexpected result is that each harmonic partial at the same time is the system's eigenfunction, which is the consequence of the TH system's special symmetry. This means that the country-level synergy is formed by the summary contribution of county-level synergies, which accords with the additive nature of synergy. Norway's innovation system can be presented as a geographically distributed network with nodes relating to corresponding counties.

From the technical side, the synergy value is a monotonic function of frequency. Because the frequency value is a proxy of the speed of change of the corresponding frequency-related transmission part (otherwise, a proxy of volatility) – one can expect frequency-related synergy volatility growth with respect to its value. This can refer both to cases of transmission increase and decrease, i.e., the synergy in more coherently interacting systems grows faster than that in less-coherent ones. In the case of decline, however, initially more coherent systems degrade faster.

This raises further research questions. If we extend the scale of study from the county to firm size level under that assumption that the results are the same, then the observations would be contradict Gibrat's Law for firm sizes, which states that for all firms in a given sector, the growth of a firm is independent of its size (Gibrat, 1931). Consequently, there should be no direct correspondence between the firm's growth and its innovation capacity, which is proportional to the synergy of interaction among constituent actors. The actual functional relation between the firm's size and its innovation capacity needs further investigation to complement what is already found in the literature (e.g. Freeman & Soete, 1997).

Another finding is that the relative contribution of long-term frequencies increases with the increase of synergy values (frequency shift). One can expect the synergy volatility to shift towards long-term fluctuations with synergy growth. That is, the short term oscillations are more accentuated in regions with low synergy values (i.e., in such regions, one can discern more cycles in close proximity than in regions with higher synergy). This means high-synergy counties are more "inertial" or trend-dependent than low-synergy counties, and this applies equally to periods of boost and decline.

Although our reasoning conserns inter-human communication networks with three-lateral interactions, it may also be applicable to other systems possessing the TH structure.

Ye, F., Yu, S., & Leydesdorff, L. (2013). The Triple Helix of University-Industry-Government Relations at the Country Level, and Its Dynamic Evolution under the Pressures of Globalization, *Journal of the American Society for Information Science and Technology, 64(11)*, 2317-2325.

Yeung, R.W. (2008). *Information theory and network coding*. New York, NY: Springer.

EUROSTAT a. Methodologies and Working papers, NACE Rev. 2 Statistical classification of economic activities in the European Community. http://epp.eurostat.ec.europa.eu/cache/ITY_OFFPUB/KS-RA-07-015/EN/KS-RA-07-015-EN.PDF (accessed August 25, 2014)

EUROSTAT b. http://www.ine.es/daco/daco42/clasificaciones/cnae09/estructura_en.pdf (accessed August 25, 2014)

**Appendix A**

The Hurst method is used to evaluate autocorrelations of the time series. It was first introduced by Hurst (1951) and was later widely used in fractal geometry (Feder, 1988). The essence of the method is as follows (Quan, Rasheed, 2004, p.2004):

For a given time series $(T_1, T_2, ... T_N)$, in our case, yearly ternary transmissions for a given time period, one can consistently perform the following steps:

a) calculate the mean $m$

$$m = \frac{1}{N}\sum_{i=1}^{N} T_i \tag{A1}$$

b) calculate mean adjusted time series:

$$Y_t = T_t - m \tag{A2}$$

c) form cumulative deviate time series:

$$Z_t = \sum_{i=1}^{t} Y_i \tag{A3}$$

d) calculate range time series:

$$R_t = \max(Z_1, Z_2, ... Z_t) - \min(Z_1, Z_2, ... Z_t) \tag{A4}$$

e) calculate standard deviation time series:

$$S_t = \sqrt{\frac{1}{t}\sum_{i=1}^{t}(T_i - \bar{T}_t)^2} \tag{A5}$$

where

$$\bar{T}_t = \frac{1}{t}\sum_{i=1}^{t} T_i \tag{A6}$$

f) calculate rescaled range time series

$$(R/S)_t = \frac{R_t}{S_t} \tag{A7}$$

in expressions (A2) - (A7) $t=1,2…N$. Under the supposition that

$$(R/S)_t = Ct^H \qquad (A8)$$

The Hurst exponent $H$ can be calculated by rescaled range (R/S) analysis and defined as linear regression slope of $R/S$ vs. $t$ in log-log scale. In our case H=0.0655 (Fig. A1).

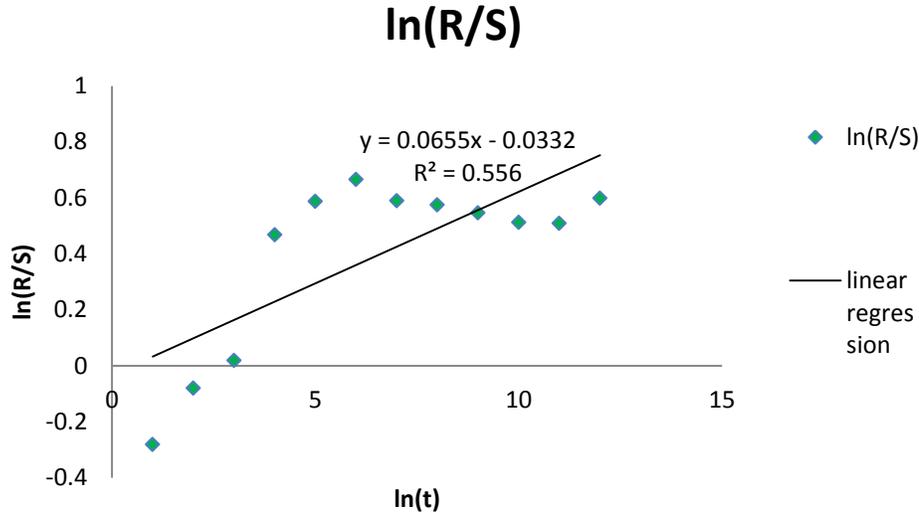

Fig. A1 R/S analysis for Norwegian synergy from 2002 to 2014

Values of H = 0.5 indicate a random time series, such as Brownian noise. Values in the interval 0 < H < 0.5 indicate anti-persistent time series in which high values are likely to be followed by low values. This tendency is more pronounced the closer the value of H comes to zero. That is, one can expect oscillating behavior. Values in the interval 0.5 < H < 1 indicate persistent time series. That is, the time series is likely to be monotonically increasing or decreasing. The case H=0.0655 corresponds to oscillatory behavior.

**Appendix B**

Table 1 Correspondence of high level aggregation to NACE Rev 1.1 and NACE Rev. 2 classifications (http://epp.eurostat.ec.europa.eu/cache/ITY_OFFPUB/KS-RA-07-015/EN/KS-RA-07-015-EN.PDF; http://www.ine.es/daco/daco42/clasificaciones/cnae09/estructura_en.pdf)

| High level aggregation | *NACE Rev.2* | *NACE Rev.1.1* |
|---|---|---|
| **1** <br><br> 1-5; <br><br> 74.14; 92.72 | **A** 1, 2, 5; Agriculture, forestry and fishing <br> 1; 2; 5; <br> 74.14; 92.72; | **A** 01 Agriculture, hunting and related service activities <br> **A** 02 Forestry, logging and related service activities <br> **A** 05 Fishing, fish farming and related service activities |
| **2** <br><br> 10-41; <br><br> 01.13; 01.41; 02.01; 51.31; 51.34; 52.74; 72.50; 90.01; 90.02; 90.03 | **B** 10-14 Mining and quarrying <br> 10 -14 | |
| | **C** 15-37 Manufacture <br> 15 - 36; <br> 01.13; 01.41; 02.01; 10.10; 10.20; 10.30; 51.31; 51.34; 52.74; 72.50; | |
| | **D** 40 Electricity, gas and steam <br> 40; | **B** 10 Mining of coal and lignite, extraction of peat <br> **B** 11 Extraction of crude petroleum and natural gas, service activities incidental to oil and gas etc. <br> **B** 12 Mining of uranium and thorium ores <br> **B** 13 Mining of metal ores <br> **B** 14 Other mining and quarrying |
| | **E (+4)** 41 Water supply, sewerage, waste <br> 41; 37; 90 <br> 14.40; 23.30; 24.15; 37.10; 37.20; 40.11; 90.01; 90.02; 90.03 | |
| **3** 45; <br> 20.30; 25.23; 28.11; 28.12; 29.22; 70.11; | **F** 45 Construction <br> 45; <br> 20.30; 25.23; 28.11; 28.12; 29.22; 70.11; | |
| **4** 50-63; <br><br> 11.10; 64.11; 64.12; | **G** 50-52 Wholesale and retail trade: repair of motor vehicles and motorcycles <br> 50- 52; | |
| | **H** 60-63 Transportation and storage <br> 60- 63; <br> 11.10; 50.20; 64.11; 64.12; | |
| | **I** 55 Accommodation and food service activities <br> 55; | **C** 15 Manufacture of food products and beverages <br> **C** 16 Manufacture of tobacco <br> **C** 17 Manufacture of textiles |
| **5** 64, 72; <br> 22.11; 22.12; 22.13; | **J** 64,72 Information and communication | **C** 18 Manufacture of wearing |
| 22.15; 22.22; 30.02; 92.11; 92.12; 92.13; 92.20; | 64; 72; 22.11; 22.12; 22.13; 22.15; 22.22; 30.02; 92.11; 92.12; 92.13; 92.20; | apparel, dressing and dyeing of fur<br>C 19 Tanning and dressing of leather, manufacture of luggage, handbags, saddlery, harness and footwear<br>C 20 Manufacture of wood and of products of wood and cork, except furniture<br>C 21 Manufacture of pulp, paper and paper products<br>C 22 Publishing, printing and reproduction of recorded media<br>C 23 Manufacture of coke, refined petroleum products and nuclear fuel<br>C 24 Manufacture of chemicals and chemical products<br>C 25 Manufacture of rubber and plastic products<br>C 26 Manufacture of other non-metallic mineral products<br>C 27 Manufacture of basic metals<br>C 28 Manufacture of fabricated metal products, except machinery and equipment<br>C 29 Manufacture of machinery and equipment n.e.c.<br>C 30 Manufacture of office machinery and computers<br>C 31 Manufacture of electrical machinery and apparatus n.e.c.<br>C 32 Manufacture of radio, television and communication equipment and apparatus<br>C 33 Manufacture of medical, precision and optical instruments, watches and clocks<br>C 34 Manufacture of motor vehicles, trailers and semi- |
| 6   65-67;<br><br>74.15; | K 65-67   Financial and insurance activities<br>65- 67;<br>74.15; | |
| 7   70; | L 70   Real estate activities<br>70; | |
| 8   71-74;<br><br>01.41; 05.01; 45.31; 63.30; 63.40; 64.11; 70.32; 75.12; 75.13; 85.20;  90.03; 92.32; 92.34; 92.40; 92.62; 92.72; | M (+10) 71,73 Professional, scientific and technical activities<br>73; 74;<br>05.01; 63.40; 85.20; 92.40;<br>N (-2) 74   Administrative and support service activities<br>71;<br>01.41; 45.31; 63.30; 64.11; 70.32; 74.50;74.87; 75.12; 75.13; 90.03; 92.32; 92.34; 92.62; 92.72; | |
| 9   75-85;<br><br>63.22; 63.23; 74.14; 92.34; 92.62; 93.65; | O 75   Public administration and defense: compulsory social security<br>75;<br>P 80   Education<br>80;<br>63.22; 63.23; 74.14; 92.34; 92.62; 93.65;<br>Q  85, 90,91   Human health and social work activities<br>85;<br>75.21; | |
| 10   92-99;<br><br>01.50;29.32; 32.20; 36.11; 36.12; 36.14; 52.71; 52.72; 52.73; 52.74; 72.50; 75.14; 91; | R 92 Arts, entertainment and recreation<br>92;<br>75.14;<br>S (+2) 93 Other service activities<br>93; 91;<br>9?":1; 01.50;29.32; 32.20; 36.11; 36.12; 36.14; 52.71; 52.72; 52.73; 52.74; 72.50;<br>T 95 Households as employers activities | |
|  | 95; | trailers |
|  | **U** 99 Extraterritorial organizations and bodies | **C** 35 Manufacture of other transport equipment |
|  | Unspecified | **C** 36 Manufacture of furniture, manufacturing n.e.c. |
|  |  | **C** 37 Recycling |
|  |  | **D** 40 Electricity, gas, steam and hot water supply |
|  |  | **E** 41 Collection, purification and distribution of water |
|  |  | **F** 45 Construction |
|  |  | **G** 50 Sale, maintenance and repair of motor vehicles and motorcycles, retail sale of automotive fuel |
|  |  | **G** 51 Wholesale trade and commission trade, except motor vehicles and motorcycles |
|  |  | **G** 52 Retail trade, except motor vehicles and motorcycles, Repair of personal and household goods |
|  |  | **I** 55 Hotels and restaurants |
|  |  | **H** 60 Land transport, transport via pipelines |
|  |  | **H** 61 Water transport |
|  |  | **H** 62 Air transport |
|  |  | **H** 63 Supporting and auxiliary transport activities, activities of travel agencies |
|  |  | **J** 64 Post and telecommunications |
|  |  | **K** 65 Financial intermediation, except insurance and pension funding |
|  |  | **K** 66 Insurance and pension funding, except compulsory social security |
|  |  | **K** 67 Activities auxiliary to financial intermediation |
|  |  | **L** 70 Real estate activities |
|  |  | **M** 71 Renting of machinery and equipment without operator and of personal and household goods |

|  |  | **J** 72 Computers and related activities |
|  |  | **M** 73 Research and development |
|  |  | **N** 74 Other business activities |
|  |  | **O** 75 Public administration and defense, compulsory social security |
|  |  | **P** 80 Education |
|  |  | **Q** 85 Health and social work |
|  |  | **Q** 90 Sewage and refuse disposal, sanitation and similar activities |
|  |  | **Q** 91 Activities of membership organizations n.e.c. |
|  |  | **R** 92 Recreational, cultural and sporting activities |
|  |  | **S** 93 Other service activities |
|  |  | **T** 95 Activities of households with employed persons |
|  |  | **U** 99 Extra-territorial organizations and bodies |